\documentclass[twocolumn,showpacs,aps,epsfig]{revtex4}

\begin{document}

 \newcommand{\Pvarphi}{{p_{\varphi}}}
 \newcommand{\rhovarphi}{{\rho_\varphi}}

 \newcommand{\bq}{\begin{equation}}
 \newcommand{\eq}{\end{equation}}
 \newcommand{\bqn}{\begin{eqnarray}}
 \newcommand{\eqn}{\end{eqnarray}}
 \newcommand{\nb}{\nonumber}
 \newcommand{\lb}{\label}
\newcommand{\PRL}{Phys. Rev. Lett.}
\newcommand{\PL}{Phys. Lett.}
\newcommand{\PR}{Phys. Rev.}
\newcommand{\CQG}{Class. Quantum Grav.}

\title{Scalar field perturbations in Horava-Lifshitz cosmology}

\author{Anzhong Wang}

\affiliation{ GCAP-CASPER, Physics Department, Baylor University,
Waco, TX 76798-7316, USA  }

\author{David Wands}

\author{Roy Maartens}

\affiliation{  Institute of Cosmology \& Gravitation, University
of Portsmouth, Portsmouth PO1 3FX, UK}

\date{\today}

\begin{abstract}

We study perturbations of a scalar field cosmology in Horava-Lifshitz gravity, 
adopting the most general setup without detailed balance but with the 
projectability condition. We derive the generalized Klein-Gordon equation, 
which is sixth-order in spatial derivatives. Then we investigate scalar field
perturbations coupled to gravity in a flat Friedmann-Robertson-Walker background. 
In the sub-horizon regime, the metric and scalar field modes have independent 
oscillations with different frequencies and phases except in particular cases.
On super-horizon scales, the perturbations become adiabatic during slow-roll 
inflation driven by a single field, and the comoving curvature perturbation is 
constant.

\end{abstract}


\maketitle

\section{Introduction}
\renewcommand{\theequation}{1.\arabic{equation}} \setcounter{equation}{0}

The background dynamics and the generation and evolution of
perturbations during a period of inflation in the early universe,
may deviate from the standard results if general relativity
acquires significant ultra-violet (UV) corrections from a quantum
gravity theory. Horava recently proposed such a theory
\cite{Horava}, motivated by the Lifshitz theory in solid state
physics \cite{Lifshitz}. Horava-Lifshitz (HL) theory has the
interesting feature that it is non-relativistic in the UV regime,
i.e., Lorentz invariance is broken. The effective speed of light
in the theory diverges in the UV regime, which could potentially
resolve the horizon problem without invoking inflation.
Furthermore, scale-invariant super-horizon curvature perturbations
could be produced without inflation \cite{Muk,KK,Piao,YKN}.
Here we consider an HL model where primordial inflation does
occur, and we investigate the changes which HL gravity induces in
the dynamics and perturbations.

Horava assumed two conditions -- detailed balance and
projectability (though he also considered the case without
detailed balance condition) \cite{Horava}. So far most of the work
on the HL theory has abandoned the projectability condition but
maintained detailed balance \cite{Cosmos,BHs,others}. One of the
main reasons is that the resulting theory is much simpler to deal
with, giving local rather than global energy constraints. However,
breaking the projectability condition is problematic \cite{CNPS}
and gives rise to an inconsistent theory \cite{LP}. With detailed
balance, on the other hand, the scalar field is not UV stable
\cite{calcagni},
and the theory requires a non-zero negative cosmological constant
and breaks parity in the purely gravitational sector \cite{SVW}
(see also \cite{KK}). To resolve these
problems, various modifications have been proposed.
The Sotiriou-Visser-Weinfurtner (SVW) \cite{SVW}
generalization is the most general setup of the HL theory with the
projectability condition and without detailed balance.
The preferred time that breaks Lorentz invariance leads to a
reduced set of diffeomorphisms, and as a result, a spin-0 mode of
the graviton appears. This mode is potentially dangerous and may
cause strong coupling problems that could prevent the recovery of
general relativity (GR) in the IR limit \cite{CNPS,LP,Mukb,Afs}.
To address this important issue and apply the theory to cosmology,
two of the current authors studied { linear} cosmological perturbations of
the Friedmann-Robertson-Walker (FRW) model with arbitrary spatial
curvature in the SVW setup,  and showed explicitly that the spin-0
scalar mode of the graviton is stable in both the IR and the UV
regimes \cite{WM}, provided that $ 0\leq\xi\leq 2/3$, where $\xi$
is a dynamical coupling parameter. { This stability condition
has the unwanted consequence that the scalar mode is a ghost
\cite{Horava,Mukb,BPS,KA}. To tackle this problem, one may
consider the theory in the range $\xi <0$, where the sound speed $c_s^2=\xi/(2-3\xi)$ is
negative. However in the limit that the sound speed becomes small as $\xi\to0$,
one should undertake a non-linear analysis to determine whether the
strong self-coupling of the scalar mode decouples \cite{Mukb}, as in the Vainshtein mechanism
in massive gravity \cite{Vain}.

In this paper we will be interested in studying cosmological perturbations in the SVW form of HL
gravity as an example of a theory which explicitly breaks Lorentz invariance. We will investigate
how standard results for linear perturbations in a scalar field cosmology are modified and how it
may still be possible to recover some standard results in the long-wavelength or low-energy limit.
We will not consider the non-linear perturbations and consider only the linear evolution of perturbations
in a scalar field cosmology. We implicitly assume that the strong-coupling (or ghost) problem can be
addressed via this mechanism or some other approach.}
 
The general equations for perturbations of an FRW universe
were derived in \cite{WM}. The coupling of matter to HL gravity
has not been worked out yet in general, as now we no longer have
the guiding principle of Lorentz invariance. Two exceptional cases
are scalar and vector fields. Scalar fields were first
investigated in \cite{calcagni} and \cite{KK}; the latter also
studied vector fields and obtained the general couplings for both
fields (see also \cite{CH}).

In Sec.~II we
obtain the stress 3-tensor for a scalar field in any spacetime,
and then derive the generalized Klein-Gordon equation, which is
sixth-order in space derivatives. In Sec.~III we specialize to an
FRW universe. We find that in the background, the generalized
Klein-Gordon equation reduces to the standard general relativistic
form, while the gravitational field equations have the Friedmann
form after replacing the Newtonian constant $G$ by $G/(1-3\xi/2)$.
But the equations for linear perturbations are quite different,
due to higher-order curvature terms. In particular, these terms
lead to a gravitational effective anisotropic stress on small
scales \cite{WM}. In Sec.~IV, we study the curvature
perturbation, showing that on large scales and in the adiabatic
case, slow-roll leads to conservation of the curvature
perturbation. We note that the large scale evolution of the
curvature perturbation in the SVW setup was studied recently
\cite{KUY}, and the conditions under which  the curvature
perturbation is conserved were discussed, but no specific matter
fields were considered.
In Sec.~V we investigate the behavior of perturbations in the sub-
and super-Hubble regimes. In Sec.~VI, we study the coupled
evolution of the adiabatic and entropy perturbations of the scalar
field. We conclude in Sec.~VII.

\section{HL Gravity with projectability and Without Detailed Balance}

\renewcommand{\theequation}{2.\arabic{equation}} \setcounter{equation}{0}

In this section, we give a very brief introduction to HL gravity
without detailed balance, but with the projectability condition.
(For further details, see \cite{SVW,WM}.)

The dynamical variables are $N, \; N^{i}$ and $g_{ij}$, in terms
of which the metric takes the ADM form,
 \bq \lb{2.1}
ds^{2} = - N^{2}dt^{2} + g_{ij}\left(dx^{i} + N^{i}dt\right)
     \left(dx^{j} + N^{j}dt\right).
 \eq
The projectability condition requires a homogeneous lapse
function,
$N = N(t)$.
%
The total action has kinetic, potential and scalar field
contributions:
  \bqn \lb{2.4}
S = \frac{1}{16\pi G} \int dt d^{3}x N \sqrt{g}
\left({\cal{L}}_{K} - {\cal{L}}_{{ V}}+ {16\pi G} {\cal{L}}_{M}
\right),
 \eqn
where 
 \bqn \lb{2.5}
{\cal{L}}_{K} &=& K_{ij}K^{ij} - \left(1-\xi\right)  K^{2},\nb\\
{\cal{L}}_{{ V}} &=& 2\Lambda - R + {16\pi G}
\left(g_{2}R^{2} +  g_{3}  R_{ij}R^{ij}\right)\nb\\
& & + \big({16\pi G}\big)^{2} \left(g_{4}R^{3} +  g_{5}  R\;
R_{ij}R^{ij}
+   g_{6}  R^{i}_{j} R^{j}_{k} R^{k}_{i} \right)\nb\\
& & + \big({16\pi G}\big)^{2} \left[g_{7}R\nabla^{2}R +  g_{8}
\left(\nabla_{i}R_{jk}\right)
\left(\nabla^{i}R^{jk}\right)\right],\nb\\
{\cal{L}}_{M}  &=& \frac{1}{2N^{2}}\left(\dot{ \varphi} -
N^{i}\nabla_{i} \varphi\right)^{2}
 - {\cal V}\left(\varphi,\nabla_i\varphi,g_{ij}\right).
  \eqn
Here 
the covariant derivatives and Ricci and Riemann terms all refer to
the three-metric $g_{ij}$, and $K_{ij}$ is the extrinsic
curvature,
$K_{ij} = \left(- \dot{g}_{ij} + \nabla_{i}N_{j} +
\nabla_{j}N_{i}\right)/2N$.
The constants $\xi, g_{I}\, (I=2,\dots 8)$  are coupling
constants, and $\Lambda$ is the cosmological constant. In the IR
limit, all the higher-order curvature terms (with coefficients
$g_I$) drop out, and the total action reduces when $\xi = 0$ to
the Einstein-Hilbert action. The potential ${\cal V}
(\varphi,\nabla_i\varphi,g_{ij})= {\cal
V}(\varphi,(\nabla\varphi)^2, {\cal P}_n)$ is defined by
\cite{KK},
%
%
    \bqn
{\cal V}  &=& V\left( \varphi\right) + \left[{1\over2}+V_{1}
\left( \varphi\right)\right] (\nabla\varphi)^2
+  V_{2}\left( \varphi\right){\cal{P}}_{1}^{2}\nb\\
& & +  V_{3}\left( \varphi\right){\cal{P}}_{1}^{3}  +
V_{4}\left( \varphi\right){\cal{P}}_{2} \nb\\
& & + V_{5}\left( \varphi\right)(\nabla\varphi)^2{\cal{P}}_{2} +
V_{6}\left( \varphi\right){{\cal P}}_{1} {\cal{P}}_{2},\nb\\
\lb{3.3}
{\cal{P}}_{n} &\equiv& \nabla^{2n} \varphi, \;\; \nabla^2 \equiv
g^{ij}\nabla_{i}\nabla_{j},
   \eqn
where $V_{s}( \varphi)$ are arbitrary functions of $ \varphi$
only. In the GR limit, $V(\varphi)$ is the usual potential, and
$V_s=0$. In order to have the scalar field stable in the UV, we
require that $V_{6} < 0$. It should be noted that the potential
(\ref{3.3}) is slightly different from the one introduced in
\cite{KK}, but they differ only by boundary terms which do not
affect the field equations.

Variation with respect to the lapse function $N(t)$ yields the
Hamiltonian constraint,
 \bq \lb{eq1}
\int{ d^{3}x\sqrt{g}\left({\cal{L}}_{K} + {\cal{L}}_{{ V}}\right)}
= 8\pi G \int d^{3}x {\sqrt{g}\, J^{t}},
 \eq
where
 \bqn \lb{eq1a}
J^{t} 
& =& - 2\left\{\frac{1}{2N^{2}} \left(\dot{ \varphi} -
N^{i}\nabla_{i} \varphi\right)^{2}
 + {\cal V} \right\}.
  \eqn
Note that, unlike GR, there is no local Hamiltonian constraint.
Variation with respect to the shift $N^{i}$ yields the
super-momentum constraint,
 \bq \lb{eq2}
\nabla_{j}\pi^{ij} = 8\pi G J^{i},
 \eq
where the super-momentum $\pi^{ij} $ and matter current $J^{i}$
are
 \bqn \lb{eq2a}
\pi^{ij} &\equiv& \frac{\delta{\cal{L}}_{K}}{\delta\dot{g}_{ij}}
 = - K^{ij} + \left(1 - \xi\right) K g^{ij},\nb\\
J^{i} &\equiv& - N\frac{\delta{\cal{L}}_{M}}{\delta N_{i}}  =
\frac{1}{N} \left(\dot{ \varphi} - N^{k}\nabla_{k}
\varphi\right)\nabla_{i} \varphi.
   \eqn

The
matter field   satisfies the conservation laws \cite{CNPS,WM},
 \bqn \lb{eq4a} & &
 \int d^{3}x \sqrt{g} { \left[ \dot{g}_{kl}\tau^{kl} -
 \frac{1}{\sqrt{g}}\left(\sqrt{g}J^{t}\right)^{\displaystyle{\cdot}}
  \right.  }   \nb\\
 & &  \left.  \;\;\;\;\;\;\;\;\;\;\;\;\;\;\;\;\;\; +  \frac{2N_{k}}
 {N\sqrt{g}}\left(\sqrt{g}J^{k}\right)^{\displaystyle{\cdot}}
 \right] = 0,\\
\lb{eq4b} & & \nabla^{k}\tau_{ik} -
\frac{1}{N\sqrt{g}}\left(\sqrt{g}J_{i}
\right)^{\displaystyle{\cdot}} - \frac{N_{i}}{N}\nabla_{k}J^{k} \nb\\
& & \;\;\;\;\;\;\;\;\;\;\; - \frac{J^{k}}{N}\left(\nabla_{k}N_{i}
- \nabla_{i}N_{k}\right) =
 0.
\eqn

Varying the action with respect to $g_{ij}$ leads to the dynamical
equations,
 \bqn \lb{eq3}
&&
\frac{1}{N\sqrt{g}}\left(\sqrt{g}\pi^{ij}\right)^{\displaystyle{\cdot}}
= -2K^i_kK^{kj}+2 \left(1 - \xi\right)K K^{ij}
\nb\\
& &~~ + \frac{1}{N}\nabla_k\left[N^k \pi^{ij}-2\pi^{k(i}
\nabla_kN^{j)}\right] \nb\\
& &~~ + \frac{1}{2} {\cal{L}}_{K}g^{ij}   + F^{ij} + 8\pi G
\tau^{ij},
 \eqn
where $F_{ij}$ is given in the Appendix.

The stress 3-tensor $\tau_{ij}$ for a scalar field is given by
 \bqn \label{tau}
 \tau_{ij} &\equiv&  -  {2\over \sqrt{g}}{\delta \left(\sqrt{g}
 {\cal{L}}_{M}\right)\over \delta{g}^{ij}} \nb\\
&&~ = {\cal{L}}_{M} g_{ij} + \left(\nabla_{i} \varphi\right)
 \left(\nabla_{j} \varphi\right) (1+ 2V_1+2V_5 {\cal P}_2)  \nb\\
& &~ +   g_{ij} \left(\nabla^2 \varphi\right){\cal V}_{,1} +
\left(\nabla^{k}
{\cal V}_{,1}\right)\left(\nabla_{k} \varphi\right) g_{ij} \nb\\
& &~ - 2\left(\nabla_{(i}{\cal V}_{,1}\right) \left(\nabla_{j)}
\varphi\right)
+ g_{ij}\left(\nabla^{4} \varphi\right){\cal V}_{,2} \nb\\
& &~ - 2 \nabla_{(i}\nabla_{k}\big[\left(\nabla^{k}{\cal
V}_{,2}\right)
\left(\nabla_{j)} \varphi\right)\big] \nb\\
& &~
 - 2 \left(\nabla^{k}  {\cal V}_{,2}\right) \left(\nabla_{k}
\nabla_{i}\nabla_{j} \varphi\right) \nb\\&&~
 +    g_{ij}   \left(\nabla^{k} {\cal V}_{,2}\right)
\left(\nabla_{k}\nabla^2 \varphi\right)  \nb\\
& &~
 - 2\left(\nabla_{(i}{\cal V}_{,2}\right) \left(\nabla_{j)}
\nabla^2 \varphi\right) \nb\\ &&~
    + 2\nabla_{k}\big[\left(\nabla^{k}{\cal V}_{,2}\right)
\left(\nabla_{i}\nabla_{j} \varphi\right)\big] \nb\\
& &~
 + 2\nabla_{(i}\big[\left(\nabla^{k}{\cal V}_{,2}\right)
\left(\nabla_{k}\nabla_{j)} \varphi\right)\big] \nb\\ &&~
 - g_{ij}\nabla^{k}\big[\left(\nabla^{l}{\cal V}_{,2}\right)
\left(\nabla_{l}\nabla_{k} \varphi\right)\big]\nb\\& &~
 + g_{ij}\nabla_{k}\nabla_{l}\big[ \left(\nabla^{k}
\varphi\right)\left(\nabla^{l}{\cal V}_{,2}\right)\big].
  \eqn

Variation of the total action with respect to $ \varphi$ yields
the generalized Klein-Gordon equation,
  \bqn
   \lb{3.4}
& & \frac{1}{N\sqrt{g}} \left[\frac{\sqrt{g}}{N}\left(\dot{
\varphi} - N^{i}\nabla_{i}
\varphi\right)\right]^{\displaystyle{\cdot}}
   = \nabla_{i}\left[\frac{N^{i}}{N^{2}}\left(\dot{ \varphi} -
   N^{k}\nabla_{k} \varphi\right)\right]\nb\\
 & & ~~~~~{} +  \nabla^{i}\left[\nabla_{i} \varphi
 \left( 1+2V_1+2V_5 {\cal P}_2 \right)\right] \nb\\ &&
  ~~~~~{}  -  {\cal V}_{, \varphi} - \nabla^2\left({\cal
V}_{,1}\right) - \nabla^{4}\left({\cal V}_{,2}\right),
  \eqn
where
  \bqn
 \lb{3.4a}
{\cal V}_{, \varphi} &\equiv& \frac{\partial {\cal V}}{\partial
\varphi} =  V'  + V_{1}' (\nabla\varphi)^2
   +  V_{2}'{\cal{P}}_{1}^{2}  \nb\\
  & &  +  V_{3}'{\cal{P}}_{1}^{3}  + V_{4}'{\cal{P}}_{2} + V_{5}'
  (\nabla\varphi)^2{\cal{P}}_{2} + V_{6}'{{\cal P}}_{1} {\cal{P}}_{2},\nb\\
{\cal V}_{,1} &\equiv& \frac{\partial {\cal V}}{\partial
{\cal{P}}_{1}}
    =  2V_{2} {\cal{P}}_{1} + 3 V_{3} {\cal{P}}_{1}^{2}
    + V_{6} {\cal{P}}_{2},\nb\\
{\cal V}_{,2} &\equiv& \frac{\partial {\cal V}}{\partial
{\cal{P}}_{2}}
    =   V_{4} + V_{5} (\nabla\varphi)^2 + V_{6} {\cal{P}}_{1}.
 \eqn

\section{Cosmological Perturbations in a Flat FRW Background}

\renewcommand{\theequation}{3.\arabic{equation}} \setcounter{equation}{0}

For the homogeneous and isotropic  FRW universe with scale factor
$a(\eta)$ and conformal Hubble rate ${\cal H}=a'/a$, the
gravitational field equations, coupled with a scalar field
described in the last section, are given by Eqs.~(\ref{A.1a}) and
(\ref{A.1b}) in the appendix. In the flat case, they reduce to
   \bqn
  \lb{4.1a}
&& \left(1 - \frac{3}{2}\xi\right)\frac{{\cal{H}}^{2}}{a^{2}}
= \frac{8\pi G}{3}\bar \rho_{ \varphi} + \frac{\Lambda}{3},\\
\lb{4.1b} && \left(1 -
\frac{3}{2}\xi\right)\frac{{\cal{H}}'}{a^{2}} =  - {4\pi G\over
3}(\bar\rho_{ \varphi}+3\bar p_{ \varphi})+ {1\over3} \Lambda,
 \eqn
where
 \bqn \lb{4.1ca}
 \bar \rho_{ \varphi} = \frac{1}{2a^{2}}{ \bar{\varphi}}^{\prime2}
 + V( \bar{\varphi}),~~
\bar p_{ \varphi}   = \frac{1}{2a^{2}}{ \bar{\varphi}}^{\prime2} -
V( \bar{\varphi}),
 \eqn
From Eqs.(\ref{4.1a}) - (\ref{4.1ca}), or directly from
Eq.~(\ref{3.4}), we find
 \bq \lb{4.1c}
\bar{\varphi}'' + 2{\cal{H}}{ \bar{\varphi}'} + a^{2}V'(
\bar{\varphi}) = 0,
 \eq
thus recovering the standard Klein-Gordon equation. All the
corrections due to high-order curvature terms vanish, and Eqs.
(\ref{4.1a})--(\ref{4.1c}) are identical to those in GR, with
modified effective gravitational constant $G \to
G_{eff}=G/(1-3\xi/2)$. Therefore, all the results obtained for
scalar field cosmologies in GR for a spatially-flat FRW background
are equally applicable to the spatially flat HL universe,
including those for inflation, as far as only the homogeneous
background is concerned. For example, the conditions for slow-roll
inflation in the flat HL universe are
$\epsilon_{V}, |\eta_{V}| \ll 1$,
where
 \bq \lb{4.1e}
\epsilon_{V} \equiv \frac{1-3\xi/2}{16\pi G}\, \frac{V^{\prime 2}
}{V^2},\;\;\; \eta_{V} \equiv \frac{1-3\xi/2}{8\pi G}\,
\frac{V''}{V}. \eq

However, inhomogeneous perturbations will be quite different, as
the higher-order curvature corrections now have non-zero
contributions.
In the quasi-longitudinal gauge \cite{WM}
 \bq ds^2 = a^2
\left[ -d\eta^2 + 2B_{,i}dx^id\eta + (1-2\psi)d\vec x\,^2 \right]
,
 \eq
  we find that
 \bqn
\lb{4.1f} && J^{t}
= - 2\left(\bar{\rho}_{\varphi}
+ \delta\rhovarphi\right),
~~ J_{i}
 = \partial_i q_\varphi,\nb\\
&& \tau^i_j
=\frac{1}{a^{2}}\Big[\left(\bar{p}_{\varphi}+\delta{\Pvarphi} +
2\bar{p}_{\varphi} \psi\right)\delta^i_j \nb\\
&&~~~~~~~ +\left(\partial^i\partial_j-{1\over3}\delta^i_j\nabla^2
\right) \Pi_\varphi\Big],
 \eqn
where
  \bqn
 \lb{4.1g}
\delta\rhovarphi &=&
\delta\rho_\varphi^{GR}+\frac{V_{4}}{a^{4}}\nabla^{4}\delta
\varphi= \frac{ \bar{\varphi}'}{a^{2}}\delta \varphi'
+ V' \delta \varphi + \frac{V_{4}}{a^{4}}\nabla^{4}\delta \varphi,\nb\\
\delta{\Pvarphi} &=&\delta p_\varphi^{GR}= \frac{1}{a^{2}}\left(
\bar{\varphi}'
\delta \varphi' - a^{2}V' \delta \varphi\right),\nb\\
q_\varphi &=&q_\varphi^{GR}= \frac{ \bar{\varphi}'}{a}\delta
\varphi=-a(\bar\rho_\varphi+ \bar p_\varphi)v_\varphi,~
v_\varphi=-{\delta\varphi' \over \bar\varphi'}\nb\\
\Pi_\varphi &=& \Pi_\varphi^{GR}= 0.
 \eqn

The linearization of the generalized Klein-Gordon equation
(\ref{3.4}) yields
  \bqn
  \lb{4.2g}
& & \delta \varphi'' + 2 {\cal H}\delta \varphi' + a^{2}V'' \delta
\varphi -\nabla^2\delta\varphi- \bar{\varphi}'\left(3\psi'
+ \nabla^{2}B\right)\nb\\
 & & ~~ =  2\Bigg({V_{1}}
     - \frac{V_{2} + V_{4}'}{a^{2}} \nabla^{2}
- \frac{V_{6}}{a^{4}} \nabla^{4}\Bigg) \nabla^{2}\delta
\varphi,
  \eqn
where the deviations from GR are on the right, and are gradient
terms, as expected. From Eqs.~(\ref{A.3a})--(\ref{A.6a}), we find
that for a spatially flat background the linearized Hamiltonian
constraint, energy conservation, trace dynamical equation,
super-momentum constraint, and trace-free dynamical equation are,
respectively,
  \bqn
   \lb{4.2a}
& & \int d^{3}x\Bigg[\nabla^{2}\psi -\left(1- \frac{3}{2}\xi
\right) {\cal H}
   \left(\nabla^{2}B + 3 \psi'\right)   \nb\\
& & ~~~~~ -  4\pi G  \Bigg( \bar{\varphi}'\delta \varphi' +
a^{2}V' \delta \varphi +
\frac{V_{4}}{a^{2}}\nabla^{4}\delta\varphi\Bigg)\Bigg]=0,
   ~~~~~~~~\\
 \lb{4.2b}
& & \int  d^{3}x {a^{2}\bar{\varphi}' }\Big(\delta \varphi'' + 2
{\cal H}\delta \varphi' + a^{2}V''\delta \varphi - 3
\bar{\varphi}'\psi'\Big)
\nb\\
 & & ~~ =  - \int d^{3}x\Bigg[{V_{4}} \nabla^{2}\delta\varphi' +
 \left(V_{4}'\bar{\varphi}' - V_{4}{\cal{H}}\right)\nabla^{4}
 \delta\varphi\Bigg], \\
   \lb{4.2f}
& &  \psi'' + 2{\cal{H}}\psi' - {\xi \over (2-3\xi)}\left(1 +
\frac{\alpha_{1}} {a^{2}}\nabla^{2}
+ \frac{\alpha_{2}} {a^{4}}\nabla^{4}\right)\nabla^{2}\psi \nb\\
& & ~~~~~~~~~~~~~~~ =   { 8\pi G\over (2-3\xi)}
\left( \bar{\varphi}' \delta \varphi' - a^{2}V' \delta \varphi\right),\\
\lb{4.2c} & & (2-3\xi)\psi'   - \xi \nabla^{2}B = 8\pi G
\bar{\varphi}'
\delta{ \varphi},\\
\lb{4.2d} & & \left(a^{2}B\right)' = \left(a^{2} +
{\alpha_{1}}\nabla^{2} + \frac{\alpha_{2}}
{a^{2}}\nabla^{4}\right)\psi,
  \eqn
where the HL constants $\alpha_1$, $\alpha_2$ are defined by Eq.~(\ref{alpha}).

The (global) energy conservation law, Eq.~(\ref{4.2b}), is
satisfied automatically, provided that $\delta\varphi$ satisfies
the generalized Klein-Gordon equation (\ref{4.2g}). Note also
that Eq.~(\ref{4.2f}) is not independent and can be obtained from
Eqs.~(\ref{4.2g}), (\ref{4.2c}) and (\ref{4.2d}). Therefore we are
left with three independent equations~(\ref{4.2g}), (\ref{4.2c})
and (\ref{4.2d}), and one  constraint, Eq.~(\ref{4.2a}), for the
three unknowns, $\psi, B$ and $\delta{ \varphi}$.

Equation (\ref{4.2d}) can be written as \cite{WM},
 \bq \lb{4.2da}
\Phi - \Psi    = \frac{1}{a^{2}}\left({\alpha_{1}} +
\frac{\alpha_{2}} {a^{2}}\nabla^{2}\right)\nabla^2\psi,
  \eq
where $\Phi$ and $\Psi$ are the usual gauge-invariant metric
perturbations \cite{Mukhanov}, and in the quasi-longitudinal gauge
\cite{WM} are given by
 \bq \lb{4.2h}
\Phi \equiv  {\cal{H}}B + B',\;\;\; \Psi  \equiv \psi -
{\cal{H}}B.
 \eq
It follows from Eq.~(\ref{4.2da}) that the higher-order curvature
corrections in the HL theory effectively create an anisotropic
stress \cite{WM}, $\Pi^{HL}=-(8\pi Ga^6)^{-1}(\alpha_1a^2+\alpha_2
\nabla^2) \nabla^2\psi $. On large scales this is negligible, but
on small scales it could produce significant deviations from GR.

In the GR limit, i.e., $\xi = 0= V_{s}$, these equations
reduce, respectively, to the corresponding equations given in GR
\cite{MW09}.

\section{Energy conservation and the curvature perturbation}
\renewcommand{\theequation}{4.\arabic{equation}} \setcounter{equation}{0}

An important quantity is the gauge-invariant curvature
perturbation on uniform-density hypersurfaces \cite{ MW09},
 \bq
 \lb{A.2c}
\zeta \equiv  - \psi -
\frac{{\cal{H}}}{\bar{\rho}_{\varphi}'}\delta\rhovarphi,
 \eq
which is constant on large scales for adiabatic perturbations in
GR. In GR this follows directly from local energy conservation
\cite{WMLL}. Moreover for a single scalar field in GR, the local
Hamiltonian constraint equation requires the non-adiabatic
pressure perturbation to vanish on super-Hubble scales
\cite{Gordon}. In this section, we show that the curvature
perturbation is also constant on large scales during slow-roll
inflation in the HL theory, although the mechanism whereby this
arises is quite different from the GR case.

The generalized Klein-Gordon equation~(\ref{4.2g}) can be
rewritten as a perturbed energy balance equation:
 \bqn
&&\delta\rho_\varphi' + 3{\cal
H}(\delta\rho_\varphi+\delta\Pvarphi) - (\bar{\rho}_{\varphi} +
\bar{p}_{\varphi})\left[3\psi' -
\nabla^2 (v_\varphi-B)\right]\nonumber\\
&&~~~~~~~~~~~~~~~~~~~ = \left(\bar{\rho}_{\varphi} +
\bar{p}_{\varphi}\right)\delta Q^{HL},\lb{4.2i}
 \eqn
{ where
the energy non-conservation $\delta Q^{\rm HL}$ is
defined as}
 \bqn
&& \delta Q^{HL} \equiv- \frac{V_{4}}{a^{2}{\bar{\varphi}}^{\prime
2}}\nabla^{4}\delta\varphi' +
\frac{1}{\bar{\varphi}'}\Bigg[-2V_{1}
+ \frac{2V_{6}}{a^{4}}\nabla^{4}\nb\\
& &~~~~~~ + \frac{1}{a^{2}}\left(2V_{2} + V_{4}' +
{\cal{H}}\frac{V_{4}}{\bar{\varphi}'}\right)\nabla^{2}\Bigg]
\nabla^{2}\delta\varphi.
 \eqn
{ In the GR limit $\delta Q^{HL}=0$ and we recover the standard
equation \cite{MW09}.
Non-zero terms on the right-hand-side represent the violation of local energy
conservation. We see that in HL gravity $\delta Q^{HL}$ is suppressed on
large scales, but on small scales local energy conservation is violated by higher-order
(Planck suppressed) terms.}

The curvature perturbation (\ref{A.2c}) obeys the evolution
equation,
 \bq
 \lb{A.2d}
\zeta' = - \frac{\cal{H}}{\bar{\rho}_{\varphi} +
\bar{p}_{\varphi}}\delta p_{\varphi\,nad}
 - \frac{1}{3}\nabla^{2}\left(v_\varphi -B \right)
  +\frac13 \delta Q^{HL} \,.
 \eq
The non-adiabatic pressure perturbation is
 \bq
 \lb{A.2a}
\delta p_{\varphi\,nad} \equiv   \delta{\Pvarphi} -
\frac{\bar{p}_{\varphi}'}
{\bar{\rho}_{\varphi}'}\delta\rho_\varphi
 = \delta p_{\varphi\,nad}^{GR} + \delta p_{\varphi\,nad}^{HL},
 \eq
where
 \bqn
 \lb{A.2b}
\delta p_{\varphi\,nad}^{GR} &\equiv&  \frac{2}{3a^{2}}\left(2 +
\frac{\bar{\varphi}''}{{\cal{H}}\bar{\varphi}'}\right)
\big[\bar{\varphi}'\delta\varphi' - \left(\bar{\varphi}'' -
{\cal{H}}\bar{\varphi}'\right)\delta\varphi\big],\nb\\
\delta p_{\varphi\,nad}^{HL} &\equiv&  \left(1 +
\frac{2\bar{\varphi}''}{{\cal{H}}\bar{\varphi}'}\right)
\frac{V_{4} }{3a^{4}} \nabla^{4}\delta \varphi.
 \eqn
Thus the non-adiabatic pressure perturbation has a
contribution of the same form as in GR, and a specific HL
contribution, which is negligible on large scales, but significant
on small scales. In GR we can use the Hamiltonian constraint to
show that $\delta p_{\varphi\,nad}^{GR}$ must vanish for a single
field on large scales
{ (if the curvature perturbation $\Psi$ remains finite)}.
In HL gravity of SVW form, we no longer have a local Hamiltonian
constraint. However, in the case of slow-roll inflation (or any
overdamped solution for a single scalar field) the existence of a
unique attractor solution for the scalar field ensures that the
local proper time derivative of the scalar field becomes a unique
function of the local field, $\dot\varphi=f(\varphi)$. { If the
scalar field perturbations approach the same slow-roll attractor
on large scales, then $\delta\dot\varphi=
f'(\varphi)\delta\varphi=
(\ddot\varphi/\dot\varphi)\delta\varphi$. It follows
from Eq.~(\ref{A.2b})
that $\delta p_{\varphi\,nad}^{GR}=0$. Since the HL corrections are also negligible on large
scales, we expect the perturbations to be adiabatic in the
super-horizon region during single-field slow-roll inflation, as
in GR.}

Therefore, the curvature perturbation on uniform density
hypersurfaces will be constant on superhorizon scales for
slow-roll inflation, which is the same as we obtain in GR. This is
expected, because the difference between GR and the HL theory is
principally in the UV regime, where the higher-order curvature
corrections become important. On large scales, these corrections
are negligible, and we expect that both of them will give the same
results. However, the mechanism here is quite different. In GR, it
is energy conservation that ensures $\delta p_{\varphi\,nad}^{GR}
= 0$ on large scales \cite{MW09}, while here it is the slow-roll
conditions
that give $\delta p_{\varphi\,nad}^{GR} \simeq 0$. In the HL
theory, the (local) conservation law of GR is replaced by its
integral form, Eq.~(\ref{4.2b}). This indicates that more
generally the perturbations need not be adiabatic and that the
curvature perturbation $\zeta$ may not be constant on superhorizon
scales in the HL cosmology in the absence of slow-roll (see also
the general discussion in \cite{KUY}).

\section{Sub- and Super-horizon Perturbations}
\renewcommand{\theequation}{5.\arabic{equation}} \setcounter{equation}{0}

Working in Fourier space, and defining
 \bq
 \lb{5.0}
u_{k} = a\delta\varphi_{k}, \;\;\; \chi_{k} = a\psi_{k},
 \eq
Equations (\ref{4.2g}), (\ref{4.2f}), (\ref{4.2c}) and
(\ref{4.2d}) lead to
 \bqn
 \lb{4.3a}
& & \chi_{k}' - {\cal{H}}\chi_{k} =  \frac{8\pi G}{2-3\xi}
\bar{\varphi}'
u_{k} -  \frac{\xi ak^{2}}{2-3\xi} B_{k},\\
\lb{4.3b} & & B_{k}' + 2 {\cal{H}}B_{k}   = \frac{1}{a}\left(1 -
\frac{\alpha_{1}}{a^{2}} k^{2}
+ \frac{\alpha_{2}} {a^{4}}k^{4}\right)\chi_{k},\\
  \lb{4.3c}
& & u_{k}'' + \bigg(\omega^{2}_{{\varphi}} -
\frac{a''}{a}\bigg)u_{k} =
 \bar{\varphi}'\bigg[3\big(\chi_{k}' - {\cal{H}}\chi_{k}\big) \nb\\
 & & ~~~~~~~~~~~~~~~~~~~~~~~~~~~~~~~~~~~ - k^{2}aB_{k}\bigg],\\
   \lb{4.3d}
&& \chi_{k}'' + \bigg(\omega^{2}_{\psi} -
\frac{a''}{a}\bigg)\chi_{k} =   { 8\pi G\over 2-3\xi}
\bigg[\bar{\varphi}' u_{k}'\nb\\
 & & ~~~~~~~~~~~~~~~~~~~~~~~  - \big({\cal{H}}\bar{\varphi}' +
a^{2}V'\big)u_{k}\bigg],
 \eqn
 where
 \bqn
 \lb{4.3e}
& & \omega^{2}_{\varphi} =
a^{2} V''+ k^{2}\bigg( 1+2V_{1} +
\frac{2(V_{2} + V_{4}')}{a^{2}} k^{2}-\frac{2V_{6}}{a^{4}} k^{4}\bigg)
 ,\nb\\
& & \omega^{2}_{\psi} = \frac{\xi k^{2}}{2-3\xi}
\bigg(
1- \frac{\alpha_{1}}{a^{2}}
k^{2} + \frac{\alpha_{2}}{a^{4}} k^{4}
\bigg).
 \eqn
From Eqs. (\ref{4.3c}) and (\ref{4.3e}) we can see that in order
for the scalar field to be stable in the UV regime, we require
that
$V_{6} < 0$.
Similarly, the metric perturbation $\psi$ is UV stable if $\xi
\alpha_{2}/(2-3\xi)\ge 0$.

To study the above equations further, we consider them in the
sub-horizon and super-horizon regimes separately.

\subsection{Sub-horizon scales}

{ On sub-horizon scales and for sufficiently large $k^2$ the highest-order curvature terms dominate the
dynamics. From Eq. (\ref{4.3e}), assuming $V_6\neq0$ and $\xi\alpha_2\neq0$, we have
\bq
 \lb{4.4}
\omega^{2}_{\varphi} \simeq  - \frac{2V_{6}}{a^{4}}\,k^{6} ,\;\;\;
\omega^{2}_{\psi} \simeq \frac{\xi \alpha_{2}}{(2-
3\xi)a^{4}}\, k^{6},
   \eq
and then from Eqs.~(\ref{4.3c}) and (\ref{4.3d}) have the oscillating
solutions (for $\xi \neq 2/3$),}
 \bq
 \lb{4.5}
u_{k} \simeq \frac{u_{0}}{\sqrt{\omega_{\varphi}}}
e^{i\omega_{\varphi} \eta}, \;\;\;\; \chi_{k} \simeq
\frac{\chi_{0}}{\sqrt{\omega_{\psi}}} e^{i\omega_{\psi} \eta},
  \eq
where $u_{0}$ and $\chi_{0}$ are constants. As
noticed by \cite{Muk}, the dispersion relationship
(\ref{4.4}) yields scale-invariant primordial perturbations.
From Eqs.~(\ref{4.3c})--(\ref{4.3e}) one can see that the scale-invariance
is not exact \cite{YKN}, due to the low-energy corrections and the
{ coupling to} metric perturbations.

In the UV regime, the scalar field mode $u_{k}$ and the metric
perturbation mode $\chi_{k}$ are oscillating independently,
although the two  metric perturbation modes $\chi_{k}$ and $B_{k}$
are oscillating with the same frequency but a different constant
phase:
 \bq \lb{4.6}
B_k \simeq - i\chi_{0}\alpha_{2}
\bigg(\frac{2-3\xi}{\xi\alpha_{2}}\bigg)^{1/2}
\frac{k}{a^{3}\sqrt{\omega_{\psi}}}e^{i\omega_{\psi} \eta},
 \eq
which follows from Eq.~(\ref{4.3a}).



When $\xi = 0$ (which corresponds to the limit of GR in the IR
regime), or $\xi = 2/3$ (when the theory has an additional symmetry,
the anisotropic Weyl invariance \cite{Horava}),  $u_k$ is still given by Eq.~(\ref{4.5}),
but the metric modes oscillate with same frequency, $\omega_\varphi$, and so are coupled to the scalar field mode. 

\subsection{Super-horizon scales}

When $k \ll {\cal{H}}$ then, up to order $k^{2}$, Eqs.~(\ref{4.3a})--(\ref{4.3c}) become
  \bqn
 \lb{4.9a}
& & \chi_{k}' - {\cal{H}}\chi_{k} =  \frac{8\pi G}{2-3\xi}
\bar{\varphi}'
u_{k} -  \frac{\xi ak^{2}}{2-3\xi} B_{k},\\
\lb{4.9b} & & B_{k}' + 2 {\cal{H}}B_{k}   = \frac{1}{a}\left(1 -
\frac{\alpha_{1}}{a^{2}} k^{2}\right)\chi_{k},\\
  \lb{4.9c}
& & u_{k}'' + \bigg(\omega^{2}_{{\varphi}} -
\frac{a''}{a}\bigg)u_{k} =
 \bar{\varphi}'\bigg[3\big(\chi_{k}' - {\cal{H}}\chi_{k}\big) \nb\\
 & & ~~~~~~~~~~~~~~~~~~~~~~~~~~~~~~~~~~~ - k^{2}aB_{k}\bigg],\\
   \lb{4.9d}
&& \chi_{k}'' + \bigg(\omega^{2}_{\psi} -
\frac{a''}{a}\bigg)\chi_{k} =   { 8\pi G\over 2-3\xi}
\bigg[\bar{\varphi}' u_{k}'\nb\\
 & & ~~~~~~~~~~~~~~~~~~~~~~~  - \big({\cal{H}}\bar{\varphi}' +
a^{2}V'\big)u_{k}\bigg],
 \eqn
where, to order $k^2$, we have from Eq.~(\ref{4.3e})
 \bq
 \lb{4.9e}
\omega^{2}_{\varphi}  \simeq  a^{2} V''+ \big(1 + 2V_{1}\big)k^{2},\;\;
\omega^{2}_{\psi} \simeq \frac{\xi}{2-3\xi} k^{2}.
 \eq

To zeroth order in $k^{2}$, from Eqs. (\ref{4.9a}) and (\ref{4.9c}) one can obtain  an equation that only involves
$u_{k}$. Once this equation is solved, from Eqs.~(\ref{4.9a}) and (\ref{4.9b}) one can find the metric perturbations 
$\chi_{k}$ and $B_{k}$ by quadrature. To order $k^2$ Eqs.~(\ref{4.9a})--(\ref{4.9c}) reduce to those in GR 
\cite{MW09} if $\alpha_{1} \to 0$ and $V_{1} \to 0$ with $G\to G_{eff}$ if $\xi\neq 2/3$, but without a local Hamiltonian 
constraint equation.



In the extreme slow-roll (de Sitter) limit,  we take $\bar{\varphi}' \simeq 0
\simeq V'$, and $a \simeq -(H\eta)^{-1}$ (with $H$ constant),
 \bqn
  \lb{4.11aa}
{\omega}_{\varphi}^{2} -
\frac{a''}{a} &=& \big(1 + 2V_{1}\big)k^{2}  - \left(1 - \frac{3\eta_{V}}{2 - 3\xi}\right) \frac{2}{\eta^{2}}. ~~~~~
 \eqn
If in addition we take the massless limit, $\eta_V\simeq0$, Eq.~(\ref{4.9c}) has the
solution to order $k^{2}$
 \bqn
 \lb{4.11ac}
 u_{k} &=& -\frac{C_{\varphi}}{H\eta} \left[ 1 + \frac{1}{2}\left( 1+ 2V_{1}\right)k^{2}\eta^2 \right]
  \nb\\
&& + D_{\varphi} \eta^2 \left[ 1- \frac{1}{10}(1+2V_1)k^2\eta^2 \right] \, .
 \eqn
The first term represents the growing mode which corresponds to a constant scalar field perturbation on large scales, 
$\delta\varphi_k\to C_\varphi$ as $k\eta\to0$, while the second term is the decaying mode. 

As in GR the scalar field perturbations decouple from the metric perturbations in the de Sitter limit. In GR this is because the local constraint equations require gauge-invariant scalar metric perturbations to vanish in this limit, but in HL gravity the metric has independent scalar perturbations.
Integrating Eq. (\ref{4.9d}) we obtain 
\bqn
\lb{4.11ad}
\chi_{k} 
 &\simeq& - \frac{C_{\chi}}{H\eta}\left[1 + \frac{1}{2}\left(\frac{\xi k^2}{2-3\xi}\right) \eta^{2}\right]\nb\\
 && + \frac{D_\chi}{3H}\left(\frac{\xi k^2}{2-3\xi}\right) \eta^2 \left[ 1- \frac{1}{10}\left(\frac{\xi k^2}{2-3\xi}\right) \eta^2 \right]. ~~~~~~~
 \eqn
Then, from Eqs. (\ref{5.0}), (\ref{4.9a}) and
 (\ref{4.9b}) we find that
 \bqn
 \lb{4.11ae}
 \psi_{k} &\simeq& C_{\chi} \left[1 + \frac{1}{2}\left(\frac{\xi k^2}{2-3\xi}\right)\eta^{2}\right],\nb\\
 && - \frac{D_\chi}{3}\left(\frac{\xi k^2}{2-3\xi}\right)\eta^3 \left[ 1 - \frac{1}{10}\left(\frac{\xi k^2}{2-3\xi}\right) \eta^2 \right] \,,\nb\\
 B_{k} &\simeq&- C_{\chi} \eta\left[ 1 - \frac{1}{2}\left(\frac{\xi}{2-3\xi} - 2\alpha_{1}H^{2} \right)k^{2}\eta^{2}\right] \nb\\
 && + D_\chi \eta^2 \left[ 1-\frac16\left(\frac{\xi k^2}{2-3\xi}\right) \eta^2 \right] \, .
 \eqn
On large scales we have $\psi_k=-B_k/\eta\to C_\chi$ as $k\eta\to0$, but this corresponds to a
gauge mode. In terms of the gauge-invariant quantities (\ref{4.2h}) we find
 \bqn
\lb{4.11af}
\Phi_{k} &\simeq&  C_\chi \left[\frac{\xi}{2-3\xi} - 2\alpha_{1}H^{2} \right]k^{2} \eta^{2} \nb\\
&& + D_\chi\eta \left[ 1 - \frac12\left(\frac{\xi k^2}{2-3\xi}\right) \eta^2 \right] \, ,\nb\\
\Psi_{k} &\simeq&  C_\chi \left[\frac{\xi}{2-3\xi} - \alpha_{1}H^{2} \right]k^{2} \eta^{2} \nb\\
&& + D_\chi\eta \left[ 1 - \frac12\left(\frac{\xi k^2}{2-3\xi}\right) \eta^2 \right] \,,
 %
 \eqn
from which we find that $\Phi_{k} - \Psi_{k}  \simeq  -C_\chi \alpha_{1}H^{2} k^{2}\eta^{2}$. 

Thus although the gauge invariant metric perturbations $\Phi$ and $\Psi$ are not constrained to vanish in the slow-roll limit, their dynamical evolution leads to $\Phi=\Psi\to0$ 
at late times  ($\eta\to0$). Similarly, although the HL theory does lead to an effective anisotropic stress (\ref{4.2da}), this is of order $k^2$ and vanishes in the large-scale limit.

\section{Coupled adiabatic and entropy perturbations on large scales}
\renewcommand{\theequation}{6.\arabic{equation}} \setcounter{equation}{0}


The gauge-invariant variable $\zeta$ is closely related to the
comoving curvature perturbation for scalar field perturbations
\cite{MW09,Mukhanov}
 \bq
{\cal R} \equiv \psi + \frac{{\cal
H}}{\bar{\varphi}'}\delta\varphi = -\zeta + {\cal H} \left(
\frac{\delta\varphi}{\varphi'} -
\frac{\delta\rho_\varphi}{\bar{\rho}'_{\varphi}} \right) \,.
 \eq
The two variables coincide, up to a choice of sign, for adiabatic
perturbations. Thus the comoving curvature perturbation should
also be conserved on large scales for adiabatic scalar field
perturbations. From the definition of the comoving curvature perturbation it is
straightforward to derive
 \bq {\cal R}' = {\cal H}{\cal S} + \psi'
+ \frac{{\cal H}'-{\cal H}^2}{\bar{\varphi}'}\delta\varphi \,,
 \eq
where we define the dimensionless intrinsic entropy perturbation
for the field
 \bq {\cal S} \equiv
{ \delta\varphi'\over \bar{\varphi}'}-{(\bar{\varphi}''-{\cal
H}\bar{\varphi}')\over \bar{\varphi}^{\prime2}}\,\delta\varphi \,.
 \eq
Note that the GR non-adiabatic pressure
perturbation in Eq.~(\ref{A.2b}) is given by
 \bq
\delta p_{\varphi\,nad}^{GR} = -
\frac{2\bar{\varphi}'V'}{3{\cal{H}}}\, {\cal S} .
 \eq

Using the HL super-momentum constraint (\ref{4.2c}),
 \bq
 \label{Reom}
{\cal R}' = {\cal H}{\cal S} + \frac{\xi}{2-3\xi} \nabla^2 B \,.
 \eq
In the GR limit, when $\xi=0$, this reduces to
${\cal R}'={\cal H}{\cal S}$ on all scales.

Using the generalized Klein-Gordon equation~(\ref{4.2g}), we obtain a first-order equation for ${\cal S}$ on
large scales
 \bqn {\cal S}' + \left(
2\frac{\bar{\varphi}''}{\bar{\varphi}'}+{\cal H} \right) {\cal S}
&=& \frac{(1+2V_1)}{\bar{\varphi}'}\nabla^2\delta\varphi \nb\\
&&~ + \frac{2}{2-3\xi} \nabla^2 B + {\cal O}(\nabla^4),~
 \eqn
where ${\cal O}(\nabla^4)$ denotes Planck-suppressed
higher-order terms.
In slow-roll, and neglecting spatial gradients on large scales, we
find
 \bq
 \label{Seom}
 {\cal S}' + 3{\cal H}{\cal S} \simeq 0,~~  {\cal R}''+2{\cal H}{\cal R}' \simeq 0 .
 \eq
Thus we find a constant mode and a rapidly decaying mode on large
scales \bq {\cal R} \simeq C + D\int \frac{d\eta}{a^2} \,. \eq
This is the same slow-roll expression as is found in GR, and is
consistent with our earlier result that $\zeta$ is conserved for
adiabatic perturbations on large scales.

However, again we see that the derivation is rather
different from GR. The local Hamiltonian constraint in GR enforces
adiabaticity on large scales \cite{MW09} \bq {\cal S} =
\frac{1}{4\pi G\bar{\varphi}'^{2}} \nabla^2 \Psi.
 \eq
In the HL case we have no such local constraint, but
slow-roll evolution (\ref{Seom}) leads to rapidly decaying entropy
perturbations at late times.

Finally, we note that in the study of  perturbations for a single scalar field in GR,
the gauge-invariant field perturbation \cite{MW09},
$\delta\varphi_{f} = \delta\varphi +
{\bar{\varphi}'}\psi/{\cal{H}} $,
is often used. The Klein-Gordon equation can be cast in a form
that involves only
$ \delta\varphi_{f}$ \cite{Hwang93}.
We find that this becomes impossible in HL gravity for two
reasons. (a)~In GR, the super-Hamiltonian constraint is used to
eliminate the metric perturbations. However, in HL theory, the
constraint is replaced by an integral form (\ref{4.2a}), which
cannot be used in the same way. (b)~Higher-order curvature
corrections enter the field equations, and these terms vanish only
on super-horizon scales.
In terms of $\delta\varphi_{f}$, the generalized Klein-Gordon equation
(\ref{4.2g}) is given in an Appendix.

\section{Conclusions}

We have studied perturbations of a scalar field cosmology in
Horava-Lifshitz gravity with projectability and without detailed balance.
After giving the field equations for an arbitrary spacetime in Sec. II, including the
generalized Klein-Gordon equation (which is sixth-order in spatial derivatives), we investigated linear perturbations about a flat
FRW universe. In the flat FRW background, the field equations and generalized Klein-Gordon equation reduce
to those in GR (under $G \to G_{eff}$). As a result, all the usual results regarding
scalar field dynamics and slow-roll inflation in the flat FRW
background also hold in the HL theory.
However, the linear perturbations
are quite different, due to the higher-order curvature terms in
the effective action which enter the equations as higher order
spatial derivatives.
In addition, the Hamiltonian constraint and the conservation of
energy now take integral forms.

In Sec. IV, we considered the evolution of $\zeta$, the curvature
perturbation on uniform-density hypersurfaces, which is
conserved for adiabatic perturbations on large scales in GR. We
identified the non-adiabatic pressure perturbation, which generalizes
the expression in GR via a
higher-order gradient correction. On large scales, the correction
vanishes, while GR part vanishes due to the slow-roll
conditions. Therefore, similar to GR,
super-horizon curvature perturbations are adiabatic and conserved (for the
curvature perturbation on uniform density hypersurfaces and the comoving curvature perturbation).
However, the mechanism for conservation is different from GR. In GR, it is the
local Hamiltonian constraint that enforces
$\delta p_{\varphi \,nad}^{GR} \simeq 0$ on large scales, while here it is the slow-roll dynamics. In the HL theory, the
conservation law of GR  is replaced by its integral form. This indicates that in more general cases than slow-roll, the scalar field
perturbations need not be adiabatic on large scales, and consequently the curvature perturbation need not be constant. This is an aspect of HL cosmology that deserves
further investigation.

In Sec.~V, we investigated the perturbations in the
sub- and super-horizon limits. In the UV sub-horizon limit, the
dispersion relations for scalar field and metric modes is of the form $\omega^2 \propto k^6$, and it has been argued that this can lead to
scale-invariant primordial perturbations \cite{Mukb}.
We identified the low-energy corrections to exact scale-invariance. The UV metric and
scalar field modes oscillate independently with
different frequencies and phases, except for the two special cases
$\xi = 0$ and $\xi = 2/3$. At these two fixed points, they are
oscillating with the same frequency, although still with different
phases. In the IR super-horizon limit, the
coupled equations reduce to a single second-order equation, and we solved for the gauge-invariant metric potentials in the de Sitter limit. In Sec.~VI we showed,
using the coupled adiabatic and entropy perturbations, how a
constant curvature perturbation is recovered on large scales in
slow-roll inflation.

\begin{acknowledgments}

AW thanks the Institute of Cosmology and Gravitation at Portsmouth for their
hospitality while the present work was initiated.    AW was
partially supported by Baylor University and the NSFC  grants,
Nos.~10703005 and 10775119. RM and DW were supported by the
UK's Science $\&$ Technology Facilities Council (STFC).
\end{acknowledgments}

\appendix
\section{The $F_{ij}$ tensor}
\renewcommand{\theequation}{A.\arabic{equation}} \setcounter{equation}{0}

The $F_{ij}$ tensor in Eq.~(\ref{eq3}) is defined in an arbitrary spacetime as
 \bqn
\lb{eq3a} F^{ij}\equiv
\frac{1}{\sqrt{g}}\frac{\delta\left(-\sqrt{g}
{\cal{L}}_{{V}}\right)}{\delta{g}_{ij}}
=\sum^{8}_{s=0}{\frac{g_{s} }{(16\pi G)^{n_{s}/2}} }\,
\left(F_{s}\right)^{ij},
 \eqn
where the additional constants are given by $g_{0} =  32\pi
G\Lambda$, $g_{1} = -1$, and $n_{s} =(2, 0, -2, -2, -4, -4, -4,
-4,-4)$. The geometric 3-tensors $ \left(F_{s}\right)_{ij}$ are:
  \bqn \lb{eq3b}
\left(F_{0}\right)_{ij} &=& - \frac{1}{2}g_{ij},\nb\\
\left(F_{1}\right)_{ij} &=& R_{ij}- \frac{1}{2}Rg_{ij},\nb\\
\left(F_{2}\right)_{ij} &=& 2\left(R_{ij} -
\nabla_{i}\nabla_{j}\right)R
-  \frac{1}{2}g_{ij} \left(R - 4\nabla^{2}\right)R,\nb\\
\left(F_{3}\right)_{ij} &=& \nabla^{2}R_{ij} - \left(\nabla_{i}
\nabla_{j} - 3R_{ij}\right)R - 4\left(R^{2}\right)_{ij}\nb\\
& & +  \frac{1}{2}g_{ij}\left( 3 R_{kl}R^{kl} + \nabla^{2}R
- 2R^{2}\right),\nb\\
\left(F_{4}\right)_{ij} &=& 3 \left(R_{ij} -
\nabla_{i}\nabla_{j}\right)R^{2}
 -  \frac{1}{2}g_{ij}\left(R  - 6 \nabla^{2}\right)R^{2},\nb\\
 \left(F_{5}\right)_{ij} &=&  \left(R_{ij} + \nabla_{i}\nabla_{j}
 \right) \left(R_{kl}R^{kl}\right)
 + 2R\left(R^{2}\right)_{ij} \nb\\
& & + \nabla^{2}\left(RR_{ij}\right) - \nabla^{k}\left[\nabla_{i}
\left(RR_{jk}\right) +\nabla_{j}\left(RR_{ik}\right)\right]\nb\\
& &  -  \frac{1}{2}g_{ij}\left[\left(R - 2 \nabla^{2}\right)
\left(R_{kl}R^{kl}\right)\right.\nb\\
& & \left.
- 2\nabla_{k}\nabla_{l}\left(RR^{kl}\right)\right],\nb\\
\left(F_{6}\right)_{ij} &=&  3\left(R^{3}\right)_{ij}  +
\frac{3}{2}
\left[\nabla^{2}\left(R^{2}\right)_{ij} \right.\nb\\
 & & \left.
 - \nabla^{k}\left(\nabla_{i}\left(R^{2}\right)_{jk} + \nabla_{j}
 \left(R^{2}\right)_{ik}\right)\right]\nb\\
 & &    -  \frac{1}{2}g_{ij}\left[R^{k}_{l}R^{l}_{m}R^{m}_{k} -
 3\nabla_{k}\nabla_{l}\left(R^{2}\right)^{kl}\right],\nb\\
 \left(F_{7}\right)_{ij} &=&  2 \nabla_{i}\nabla_{j}
 \left(\nabla^{2}R\right) - 2\left(\nabla^{2}R\right)R_{ij}\nb\\
 & &    + \left(\nabla_{i}R\right)\left(\nabla_{j}R\right)
  -  \frac{1}{2}g_{ij}\left[\left(\nabla{R}\right)^{2} +
  4 \nabla^{4}R\right],\nb\\
\left(F_{8}\right)_{ij} &=&  \nabla^{4}R_{ij} -
\nabla_{k}\left(\nabla_{i}\nabla^{2} R^{k}_{j}
                            + \nabla_{j}\nabla^{2} R^{k}_{i}
                            \right)\nb\\
& & - \left(\nabla_{i}R^{k}_{l}\right)
\left(\nabla_{j}R^{l}_{k}\right)
       - 2 \left(\nabla^{k}R^{l}_{i}\right) \left(\nabla_{k}R_{jl}
       \right)\nb\\
& &    -  \frac{1}{2}g_{ij}\left[\left(\nabla_{k}R_{lm}\right)^{2}
        -
        2\left(\nabla_{k}\nabla_{l}\nabla^{2}R^{kl}\right)\right].
 \eqn

\section{Cosmological Perturbations in an FRW Background}

\renewcommand{\theequation}{B.\arabic{equation}} \setcounter{equation}{0}

We summarize the key cosmological perturbation equations for the
FRW metric,
 $ds^{2} = a^{2}(- d\eta^{2}  + \gamma_{ij}dx^{i}dx^{j})$,
where $\gamma_{ij}=[1 + K(x^{2} + y^{2} + z^{2})/4]^{-2}
{\delta_{ij}}$, with $K = 0, \pm 1$. In the background, the
Hamiltonian constraint~(\ref{eq1}) and dynamical equation (\ref{eq3}) reduce to \cite{SVW,WM},
  \bqn \lb{A.1a}
\left(1 - \frac{3}{2}\xi\right)\frac{{\cal{H}}^{2}}{a^{2}} +
\frac{K}{a^{2}} &=&
\frac{8\pi G}{3}\bar \rho_{ \varphi} + \frac{\Lambda}{3}\nb\\
&&~+ \frac{2\beta_1K^{2}}{a^{4}} + \frac{4\beta_2K^{3}}{a^{6}},\\
\lb{A.1b}
\left(1 - \frac{3}{2}\xi\right)\frac{{\cal{H}}'}{a^{2}} &=&  -
{4\pi G\over 3}(\bar\rho_{ \varphi}+3\bar p_{ \varphi})+ {1\over3}
\Lambda
\nb\\  & &~~~
- \frac{2\beta_{1}K^{2}}{a^{4}}  - \frac{8\beta_{2}K^{3}}{a^{6}},
 \eqn
where
 \bq
\beta_1=16\pi G(3g_2+g_3),\, \beta_2=(16\pi
G)^2(9g_4+3g_5+g_6).
 \eq

Then to first-order,  using  \cite{WM},  the Hamiltonian and super-momentum constraints,
the trace and trace-free dynamical equations, and energy conservation are given,
respectively, by
   \bqn
   \lb{A.3a}
   & & \int
\sqrt{\gamma}d^{3}x\Bigg[\left(\nabla^2+3K\right)\psi -
\frac{{\cal H}(2-3\xi)}{2}
\left(\nabla^2 B + 3\psi'\right) \nb\\
& &~~~ -2 K\Big(\frac{2\beta_{1}}{a^{2}} +
\frac{6\beta_{2}K}{a^{4}} +
\frac{3g_{7}}{\zeta^{4}a^{4}}\nabla^2\Big)
\left(\nabla^2+3K\right)\psi\nb\\
& &   ~~~ - 4\pi G a^{2}\Big(\frac{ \bar{\varphi}'}{a^{2}}\delta
\varphi' + V' \delta \varphi
+ \frac{V_{4}}{a^{4}}{\nabla}^{4}\delta \varphi\Big)\Bigg]=0,\\
\lb{A.3b} & & (2-3\xi)\psi' - 2KB - \xi {\nabla}^{2}B = 8\pi G
\bar{\varphi}'\delta{ \varphi},\\
\lb{A.3c} & &  \psi'' + 2{\cal{H}}\psi'  - {\cal{F}}\psi + {1\over
3}\left(\nabla^2B'+2 {\cal H}
\nabla^2B \right)\nb\\
& & ~~~ - \frac{ \gamma^{ij}\delta{F}_{ij}}{3(2-3\xi)}    = { 8\pi
G\over (2-3\xi)}
\left( \bar{\varphi}' \delta \varphi' - a^{2}V'
\delta \varphi\right),\\
\lb{A.3d} & & \left(B'+2{\cal H}B\right)_{|\langle ij \rangle}
+\delta F_{\langle ij \rangle}= 0,\\
 \lb{A.6a}
& & \int  d^{3}x \Big(\delta \varphi'' + 2 {\cal H}\delta \varphi' + a^{2}V''\delta \varphi - 3  \bar{\varphi}'\psi'\Big) {a^{2}\bar{\varphi}' }\nb\\
 & & ~~ =  - \int d^{3}x\bigg[{V_{4}}  \vec{\nabla}^{2}\delta\varphi' +
 \left(V_{4}'\bar{\varphi}' - V_{4}{\cal{H}}\right) \vec{\nabla}^{4}\delta\varphi\bigg],  ~~~~~~~~~
 \eqn
where a vertical bar denotes the covariant derivative with respect
to $\gamma_{ij}$, and angled brackets on indices denote the
symmetric trace-free part. Here
 \bq \lb{A.4}
{\cal{F}} ={2a^2\over(2-3\xi)}\left(-\Lambda+{K \over
a^2}+{2\beta_1K^2
 \over a^4} +{12\beta_2 K^3 \over a^6 } \right),
 \eq
and $\delta{F}_{ij}$ is given by Eq. (A.1) in \cite{WM}. When $K =
0$, using \cite{WM},
 \bqn \lb{A.5}
&&\delta{F}_{ij} =  2\Lambda a^{2}\psi \delta_{ij}\nb\\ &&~~~~ - \left(1
+\frac{\alpha_{1}}{a^{2}}\nabla^{2}
+  \frac{\alpha_{2}}{a^{4}}\nabla^{4}\right)
\left(\partial_{i}\partial_{j}   -
\delta_{ij}\nabla^{2}\right)\psi ,~~~~~ \eqn
where
 \bq \label{alpha}
\alpha_1=16\pi G(8g_2+3g_3),\, \alpha_2=(16\pi G)^2(3g_8-8g_7).
 \eq

\section{Generalized Klein-Gordon equation in $\delta\varphi_{f}$}
\renewcommand{\theequation}{C.\arabic{equation}} \setcounter{equation}{0}

Using Eqs.~(\ref{4.2f})
and (\ref{4.2c}),
 \bqn
 \lb{4.5a}
&&  \delta\varphi_{f}'' + 2{\cal{H}}\delta\varphi_{f}' -
(1+2V_{1})\nabla^{2} \delta\varphi_{f}
  + a^{2}V''\delta\varphi_{f} \nb\\
& & ~~ + \frac{{\cal{H}}^{2} - {\cal{H}}'}{\bar{\varphi}'
{\cal{H}}^{2}}\Big\{\big[2\bar{\varphi}'\left({\cal{H}}'
+ {\cal{H}}^{2}\right) + 3 a^{2}{\cal{H}}V'\big] \delta\varphi_{f}   \nb\\
  & & ~~~~~~~~~~~~~~~~~~~~~
+   {\cal{H}} \bar{\varphi}'
  \delta\varphi_{f}' \Big\}\nb\\
  & & ~~ + \frac{2}{a^{2}}\Big[\left(V_{2} + V_{4}'\right)
  + \frac{V_{6}}{a^{2}}\nabla^{2}\Big]\nabla^{4}
  \delta\varphi_{f}\nb\\
  & & = \frac{ \bar{\varphi}'}{(2-3\xi){\cal{H}}}\Big\{
  \big[-2+4\xi - 2(2-3\xi)V_{1}\big] \nb\\
  & & ~~~~
  + \frac{1}{a^{2}}\big[\xi\alpha_{1} + 2(2-3\xi)\left(V_{2}
   + V_{4}'\right)\big]\nabla^{2}\nb\\
  && ~~~~  + \frac{1}{a^{4}}\big[\xi\alpha_{2} +
  2(2-3\xi) V_{6}\big]\nabla^{4}\Big\}\nabla^{2}\psi\nb\\
  & & ~~~~ +  \frac{1}{{\cal{H}}^{2}}\big[\bar{\varphi}'
  \left(4{\cal{H}}^{3} - 2 {\cal{H}}{\cal{H}}' - {\cal{H}}''\right) \nb\\
  & & ~~~~~~~~~~~~~~  + 2a^{2}\left({\cal{H}}^{2} -  {\cal{H}}'
  \right)V'\big]\psi\nb\\
  & & ~~~~ + \frac{1}{(2-3\xi){\cal{H}}^{2}}\big[(2-5\xi)
  \bar{\varphi}{\cal{H}}^{2} \nb\\
& & ~~~~~~~~~~~~~~~~~~~ - \xi\left(\bar{\varphi}{\cal{H}}' +
2a^{2}\bar{\varphi}{\cal{H}}V'\right)\big]\nabla^{2}B,\nb\\
  \eqn
where the gauge-invariant variable $\delta\varphi_{flat}$ is defined as
\cite{MW09},
\bq
\lb{C.2}
\delta\varphi_{flat} = \delta\varphi + \frac{\bar{\varphi}'}{\cal{H}}\psi.
\eq
As noted above, and unlike the case of GR, the metric variables $\psi$ and $B$ remain
in the equation.


\end{document}